# Radio Frequency Readout of Electrically Detected Magnetic Resonance in Phosphorus-Doped Silicon MOSFETs

L.H. Willems van Beveren*, H. Huebl, R.P. Starrett and A. Morello
ARC Centre of Excellence for Quantum Computer Technology, School of Physics,
The University of New South Wales, Sydney 2052, Australia
*Corresponding author: Email l.h.willemsvanbeveren@unsw.edu.au

**Abstract:** We demonstrate radio frequency (RF) readout of electrically detected magnetic resonance in phosphorus-doped silicon metal-oxide field-effect-transistors (MOSFETs), operated at liquid helium temperatures. For the first time, the Si:P hyperfine lines have been observed using radio frequency reflectometry, which is promising for high-bandwidth operation and possibly time-resolved detection of spin resonance in donor-based semiconductor devices. Here we present the effect of microwave (MW) power and MOSFET biasing conditions on the EDMR signals.

**1 Introduction:** Recently, broadband electrically detected magnetic resonance (EDMR) spectroscopy of phosphorus electron spins in silicon MOSFETs was achieved [1]. Here, an on-chip short-circuited coplanar stripline (CPS) was used to generate the oscillating magnetic field for electron-spin resonance, which simultaneously acted as gate electrode. This direct current (DC) spectroscopy allowed determination of all the *g*-factors.

**2 Radiofrequency readout:** Since then, we have developed the technology for radio frequency readout of EDMR, based on the reflection of a carrier signal from a resonant LCR circuit, in which the MOSFET is incorporated as resistive element [2], c.f. Fig. 1. Compared to conventional EDMR readout – based on DC current-to-voltage conversion and subsequent lock-in (LI) amplification – the RF technique displays a much higher sensitivity and operates at much higher speeds due to the high operational bandwidth involved. The improved resolution in the time-domain suggests using this RF technique for the observation of coherently manipulated spin states, e.g. Rabi oscillations directly in real-time. The EDMR spectrum obtained by DC and RF readout is shown in Fig. 2. The MW frequency used was 28.55 GHz, corresponding to ~1 T field (K-band ESR).

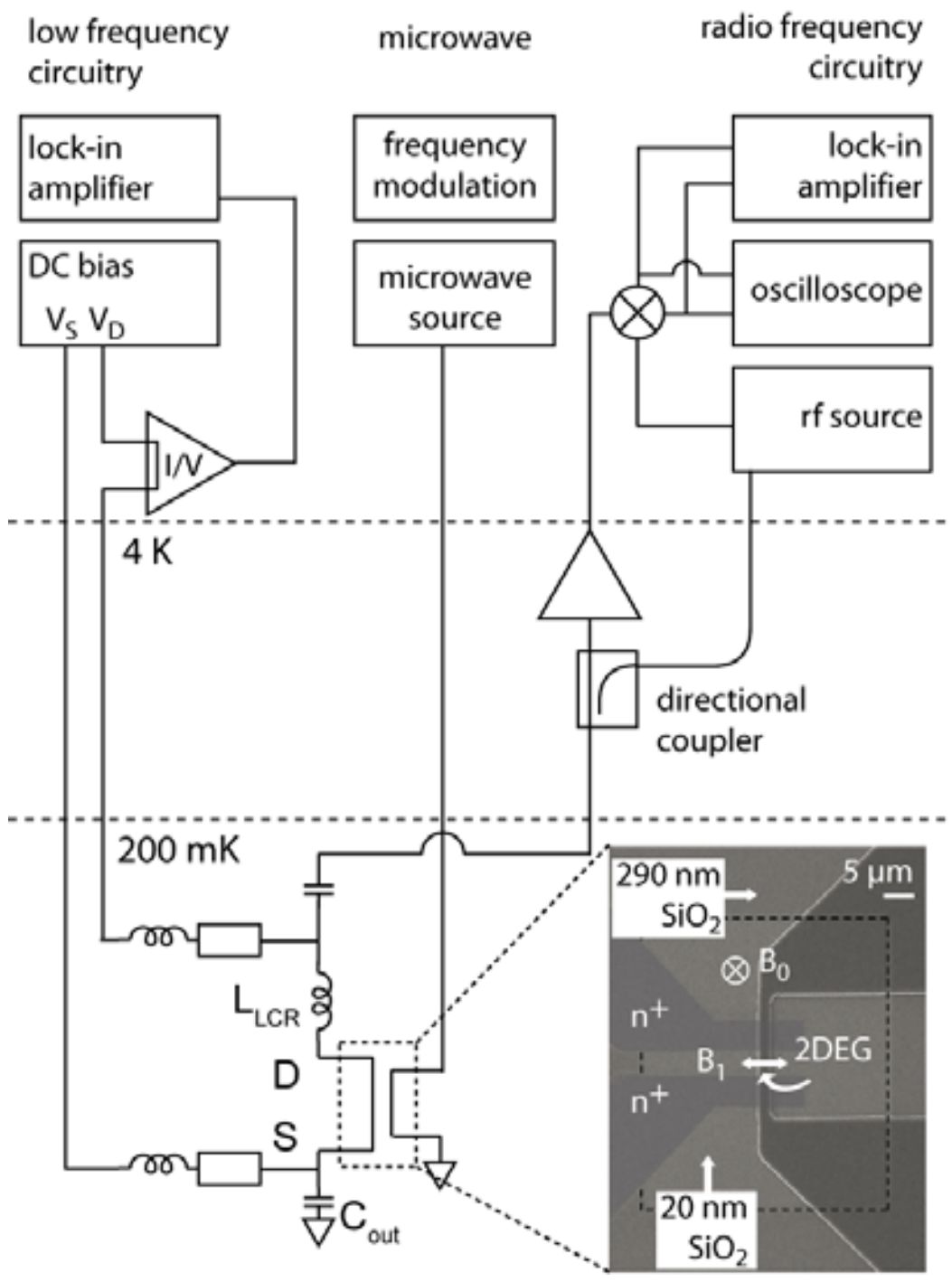

Fig.1 Low frequency (DC), microwave (MW) and radio frequency (RF) setup. Here the MOSFET is embedded as a resistive element in a low-temperature resonant LCR-circuit, allowing high-bandwidth EDMR spectroscopy.

**3 MW power dependence:** We have looked in detail at the effect of the applied microwave power and MOSFET biasing conditions on the EDMR signal intensity for both the 2DEG and Si:P hyperfine (HF) lines, following the pioneering work described in Refs. [3,4].

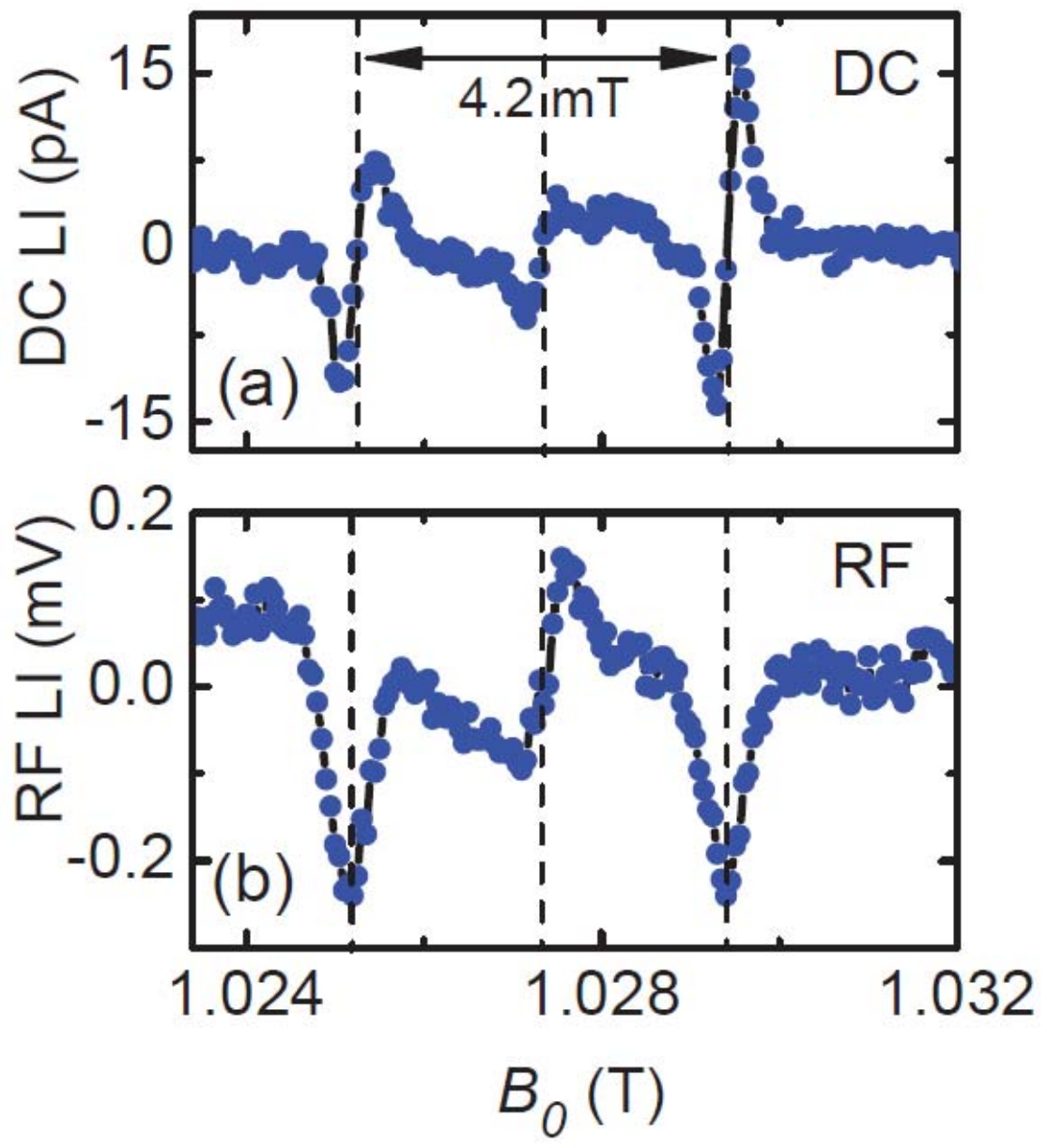

Fig.2 EDMR spectra obtained by DC (a) and RF (b) readout at a MW frequency of 28.55 GHz using frequency modulation (FM) for phase sensitive detection.

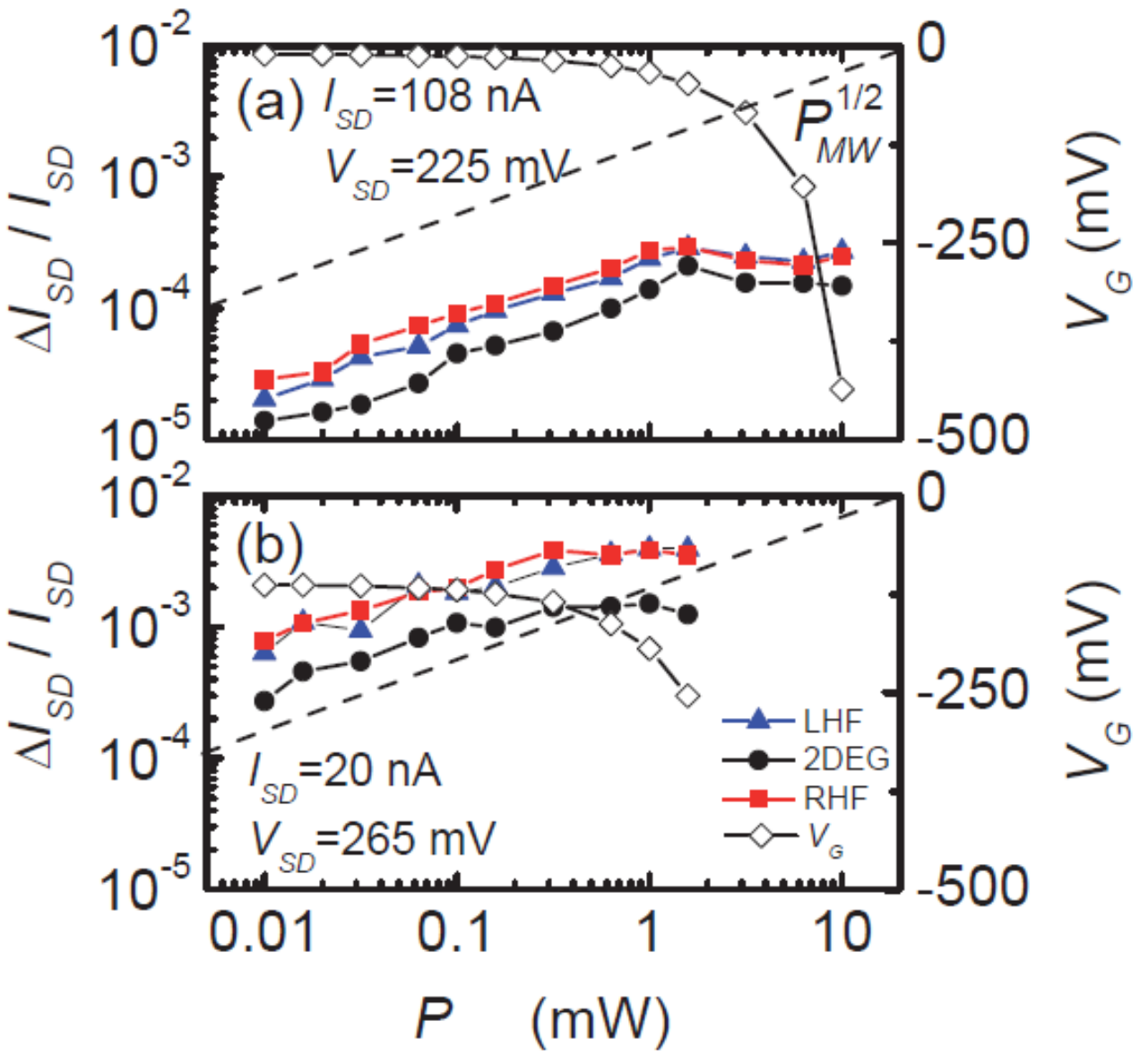

Fig.3 MW power dependence for both the 2DEG and Si:P hyperfine lines at 0.1 K for (a) $V_{SD}$=225 mV, $I_{SD}$=108 nA and (b) $V_{SD}$=265 mV, $I_{SD}$=20 nA, at 28.16 GHz.

The MW power dependence is shown in Fig. 3. The peak-to-peak signal intensity of all the EDMR resonances increases with the square root of the applied MW power, which is similar to cavity-based EPR. For sufficiently large MW powers saturation of the EPR transitions occurs.

Interestingly, the signal intensity of all resonances also depends on the source-drain voltage applied across the device. For a high bias ($V_{SD}$=265 mV) EDMR signal intensities of $\Delta I_{SD}/I_{SD}$ > $10^{-3}$ can be achieved for the Si:P donor electrons, which is in agreement with theory [5].

Finally, we observe that the saturation power depends on the value of the source-drain current. This suggests that electron scattering events determine the spin relaxation time $T_1$. Further analysis is underway to shed light on the details of the spin-dependent scattering phenomena emerging in this experiment.

## Acknowledgements

The authors would like to thank D. Barber for assistance in the National Magnet Lab. This work has been supported by the Australian Research Council (ARC), the Australian Government, the U.S. National Security Agency and the U.S. Army Research Office under Contract No. W911NF-08-1-0527.